\documentclass[aip,reprint]{revtex4-1}

\draft 

\usepackage{amssymb}
\usepackage{amsmath}
\usepackage{graphicx}
\usepackage{caption}
\usepackage{bm}
\usepackage{multirow}
\usepackage{bm}
\usepackage[colorlinks=true,linkcolor=blue,citecolor=blue,urlcolor=blue]{hyperref}

\begin{document}

\title{Runaway electron avalanche and macroscopic beam formation: simulations of the DTT full power scenario}\thanks{The following article has been submitted to Physics of Plasmas.}

\author{Enrico Emanuelli}
 \email{enrico.emanuelli@polito.it}
 \affiliation{\mbox{NEMO Group, Dipartimento Energia, Politecnico di Torino, Corso Duca degli Abruzzi 24, Torino, 10129, Italy}}
 \affiliation{\mbox{DTT S.C.a.r.l., Via Enrico Fermi 45, Frascati, 00044, Italy}}

\author{Francesco Vannini}
 \affiliation{\mbox{Max Planck Institute for Plasma Physics, Boltzmannstr. 2, Garching bei M\"unchen, 85748, Germany}}

\author{Matthias Hoelzl}
 \affiliation{\mbox{Max Planck Institute for Plasma Physics, Boltzmannstr. 2, Garching bei M\"unchen, 85748, Germany}}

\author{Eric Nardon}
 \affiliation{\mbox{CEA, IRFM, F-13108 Saint-Paul-lez-Durance, France}}

\author{Vinodh Bandaru}
 \affiliation{\mbox{Indian Institute of Technology Guwahati, North Guwahati, Assam, 781039, India}}

\author{Nina Schwarz}
 \affiliation{\mbox{ITER Organization, Route de Vinon-sur-Verdon, Saint-Paul-lez-Durance Cedex, 13067, France}}

\author{Daniele Bonfiglio}
 \affiliation{\mbox{Consorzio RFX, Corso Stati Uniti 4, Padova, 35127, Italy}}
 \affiliation{\mbox{CNR-ISTP, Corso Stati Uniti 4, Padova, 35127, Italy}}

\author{Giuseppe Ramogida}
 \affiliation{\mbox{DTT S.C.a.r.l., Via Enrico Fermi 45, Frascati, 00044, Italy}}
 \affiliation{\mbox{ENEA, Via Enrico Fermi 45, Frascati, 00044, Italy}}
 \affiliation{\mbox{Universit\`a della Tuscia, Via Santa Maria in Gradi 4, Viterbo, 01100, Italy}}

\author{Fabio Subba}
 \affiliation{\mbox{NEMO Group, Dipartimento Energia, Politecnico di Torino, Corso Duca degli Abruzzi 24, Torino, 10129, Italy}}
 
\author{JOREK Team}
 \thanks{See author list of M. Hoelzl et al 2024 Nucl. Fusion 64 112016}
 \noaffiliation

\date{December 30, 2025}

\begin{abstract}
The transition of the Divertor Tokamak Test (DTT) facility from its initial commissioning phase (Day-0, plasma current $I_{p}=2$ MA) to the full power scenario ($I_{p}=5.5$ MA) introduces a critical shift in the dynamics of runaway electrons (REs) generation. While previous predictive studies of the low-current scenario indicated a robust safety margin against RE beam formation, this work reveals that the exponential scaling of the RE avalanche gain with plasma current severely narrows the safe operational window in the full power scenario. Using the non-linear magnetohydrodynamic code JOREK, we perform comprehensive 2D simulations of the current quench (CQ) phase of several disruption scenarios, systematically scanning initial RE seed currents and injected impurity levels. The results demonstrate that in the full power scenario, the avalanche multiplication factor is sufficiently high ($G_{av} \approx 1.3 \cdot 10^5$) to convert a mere 5.5 A seed current into macroscopic RE beams of $\approx 0.7$ MA when large amounts of impurities are present. For even higher RE seeds, the RE current can peak at $ \approx 3.2$ MA, constituting up to $\approx$ 80\% of the total plasma current during the CQ. These findings suggest that, unlike the Day-0 phase, the disruption mitigation strategy for the full power scenario involves a careful balance between thermal load mitigation and RE avoidance, necessitating a well-chosen quantity of injected impurities. This work provides the baseline needed for future estimations of RE loads on the plasma-facing components of DTT, which will be essential for designing and positioning mitigation components like sacrificial limiters.
\end{abstract}

\maketitle 

\section{Introduction}
\label{sec:introduction}

The successful operation of next-generation tokamaks requires robust strategies to prevent or mitigate plasma disruptions \cite{Iter1999a, Hender2007}. Disruption mitigation typically relies on massive injection of impurities (e.g., neon or argon) to radiate the plasma thermal energy and distribute the heat load more uniformly \cite{Iter1999a, Hender2007}. This rapid cooling of the plasma (thermal quench, TQ) drastically increases the plasma resistivity, ending the discharge by leading to a fast decay of the plasma current (current quench, CQ). However, the strong induced toroidal electric field can accelerate electrons to relativistic energies, generating a population of runaway electrons (REs) \cite{Boozer2018, Breizman2019}. The primary RE population (RE seed) can be exponentially amplified via the avalanche mechanism, where relativistic electrons knock thermal electrons into the runaway region through close Coulomb collisions \cite{Rosenbluth1997}. The avalanche gain, defined as the amplification factor of the RE seed population in the limit of an extremely small initial seed, scales exponentially with the plasma current $I_p$ ($G_\text{av} \approx e^{I_p/(I_A \ln \Lambda)}$, where $I_A$ is the Alfvén current and $\ln \Lambda$ is the Coulomb logarithm), severely exacerbating the risk posed by REs in high-current regimes \cite{Rosenbluth1997}. If not suppressed, the resulting RE beam can carry a substantial fraction of the pre-disruption plasma current, eventually impinging on localized regions of the first wall and potentially causing the deep melting of metallic plasma-facing components (PFCs) \cite{Hender2007, Matthews2016, Ratynskaia2025}.

The Divertor Tokamak Test (DTT) facility \cite{Martone2019, Romanelli2024}, currently under construction, is designed to investigate power exhaust solutions in regimes relevant to ITER \cite{Iter1999b} and EU-DEMO \cite{Federici2019}. With a nominal plasma current of $I_p = 5.5$ MA and auxiliary heating power up to 45 MW, DTT will achieve divertor heat flux densities comparable to those expected in future reactors \cite{Romanelli2024}. Although the ultimate goal is operation in the full power scenario, the operational plan of DTT includes an initial low-current commissioning phase (Day-0, characterized by $I_p=2$ MA and a toroidal magnetic field of $B_\phi = 3$ T \cite{Casiraghi2023}), where the potential for RE generation is substantially lower. Previous predictive studies \cite{Emanuelli2025} focused on this Day-0 scenario of DTT and indicated that, by controlling the amount of impurities injected via the disruption mitigation system (DMS), it is possible to avoid the conditions that generate a substantial RE beam, even with large RE seeds ($\approx 20$ kA). For smaller RE seeds, the generation of RE beams remains minimal, even for high levels of impurities. However, the transition to the full power scenario ($I_p = 5.5$ MA, $B_\phi = 5.85$ T) introduces a drastic shift in disruption dynamics: since the avalanche gain $G_\text{av}$ scales exponentially with the plasma current, the 5.5 MA scenario is susceptible to the conversion of even minute seed populations into macroscopic RE beams, rendering the system more sensitive to the amount of impurities injected by the DMS. 

This work broadens the safety analysis of RE avalanches in DTT to the full power scenario, employing the non-linear magnetohydrodynamic (MHD) code JOREK \cite{Hoelzl2021, Hoelzl2024} and the resistive wall code STARWALL \cite{Merkel2015, Hoelzl2012}. Building upon the work carried out for the Day-0 scenario \cite{Emanuelli2025}, we extend the 2D analysis to the reactor-relevant regime, investigating the CQ dynamics and the formation of macroscopic RE beams at 5.5 MA. We systematically vary impurity levels and seed currents to assess the dependence of the RE beam current on these factors, clarifying the specific combinations of parameters that lead to significant RE beam formation and thereby defining the risk boundaries for the full power scenario.

\section{Models and methods}
\label{sec:model}
The computational framework employed in this study combines the non-linear MHD code JOREK \cite{Hoelzl2021, Hoelzl2024} with the resistive wall code STARWALL \cite{Merkel2015, Hoelzl2012}, closely following what was done in our previous work \cite{Emanuelli2025}. We employ the reduced MHD formulation of JOREK \cite{Hoelzl2021, Czarny2008}, where the magnetic and electric fields are expressed through the poloidal magnetic flux $\psi$ and the velocity stream function $u$. To study RE dynamics, the MHD equations are coupled with the RE fluid model described in \cite{Bandaru2019, Bandaru2024a}. This model treats REs as a distinct fluid species, self-consistently capturing the feedback between the RE population and the bulk plasma. We prescribe a pre-existing RE seed at the initialization to serve as a proxy for primary formation processes (e.g., hot-tail or Dreicer), which is then multiplied via the computation of the secondary volumetric source (RE avalanching). Previous applications of this framework include the study of RE formation and termination in JET \cite{Bandaru2021}, ITER \cite{Bandaru2024b, Bergstrom2024, Wang2024} and EU-DEMO \cite{Vannini2025}.

Following the transport coefficients detailed in \cite{Emanuelli2025}, the electrical resistivity $\eta_e$ and parallel thermal diffusivity $\chi_{\parallel}$ follow the standard Spitzer ($\propto T_e^{-3/2}$) and Spitzer-Härm ($\propto T_e^{5/2}$) scalings, respectively, but are regularized by constant cut-off values for temperatures exceeding 500 eV and 350 eV. Additionally, the parallel thermal diffusivity is subject to a low-temperature cut-off at 20 eV. Regarding cross-field transport, we employ a constant particle diffusivity $D = 2$ $\text{m}^2/\text{s}$ and a perpendicular thermal diffusivity $\chi_{\perp}$ profile with a core value of $2$ $\text{m}^2/\text{s}$. The boundary condition for the electron temperature is fixed at 0.7 eV. For the RE fluid, a dominant parallel diffusivity of $D_{\parallel, RE} = 1 \times 10^9$ $\text{m}^2/\text{s}$ is imposed to mimic the rapid transit of RE along magnetic field lines. As established in \cite{Bandaru2024a}, this specific value is optimal for the fluid model, approximating well the parallel advection without introducing significant numerical artifacts in the perpendicular direction. Conversely, the perpendicular diffusivity is set at a low value of $D_{\perp, RE} = 1 \times 10^{-2}$ $\text{m}^2/\text{s}$, which serves primarily to ensure numerical stability \cite{Bandaru2024b}. We restrict our analysis to toroidally symmetric 2D simulations, hence omitting non-axisymmetric perturbations. This yields an upper estimate of the RE beam current, as 2D simulations do not account for losses driven by stochastic field lines. Although 3D simulations are necessary to capture these loss mechanisms, their complexity and computational cost place them outside the scope of the present study. To strictly focus on CQ dynamics and RE formation while maintaining computational efficiency, we adopt a single-fluid representation with equal ion and electron temperatures ($T = T_i + T_e = 2T_e$) and neglect neutral particles, diamagnetic drifts, and parallel background fluid velocity.
While fluid variables rely on fixed boundary conditions, the poloidal flux $\psi$ and the total toroidal current density $j$ at the boundary are determined through implicit coupling with the resistive wall code STARWALL \cite{Hoelzl2012}. This allows the simulation to evolve the plasma dynamics alongside the currents induced in the passive and active structures of the DTT machine, making the simulation of the CQ phase more realistic by capturing the vertical motion of the plasma and the associated RE beam scrape-off against the PFCs. The computational boundary and the specific geometry of the DTT structures imported into STARWALL correspond to the layout detailed in Figure 1 of our previous work \cite{Emanuelli2025}.
To initialize the disruption and subsequent RE generation, we employ an artificial TQ (ATQ) mechanism \cite{Emanuelli2025, Bandaru2024b, Vannini2025, Schwarz2023, Schwarz2023b}. This method bypasses the high computational cost of full 3D TQ physics (which JOREK can simulate self-consistently \cite{Hu2023, Hu2023b, Bonfiglio2025, Nardon2021}, and will be addressed in future work) by artificially establishing post-TQ conditions, providing a controlled environment where to focus on the RE formation that occurs during the CQ phase.
The ATQ is implemented in three sequential steps, visually demarcated by the thin brown vertical lines in Figures \ref{fig:seed_6}--\ref{fig:z_seed2}. This separation into steps is strictly a numerical convenience, allowing us to address the thermal, magnetic, and impurity aspects independently. First, the perpendicular thermal diffusivity ($\chi_\perp$) is transiently increased to drive a rapid temperature collapse to $\approx 10$ eV (region preceding the brown dotted line). This is followed by the application of a high hyper-resistivity to flatten the current density profile in the core, mimicking the effects of magnetic reconnection \cite{Yamada2010} (interval between the dotted and dashed-dotted lines). Finally, neon impurities are injected uniformly at a fixed rate starting from the dashed-dotted line. Upon reaching the target density, the injection is stopped, and the resulting change in the slope of the plasma current traces in Figures \ref{fig:seed_6}--\ref{fig:z_seed2} signals the onset of the self-consistent CQ phase, where all modified parameters return to their physical values. Unlike in experiments where impurities trigger the collapse, here they are introduced to ensure a balance between radiative cooling and Ohmic heating, which is required for the self-consistent CQ.

\begin{figure}[t!]
    \centering
    \includegraphics[width=\columnwidth]{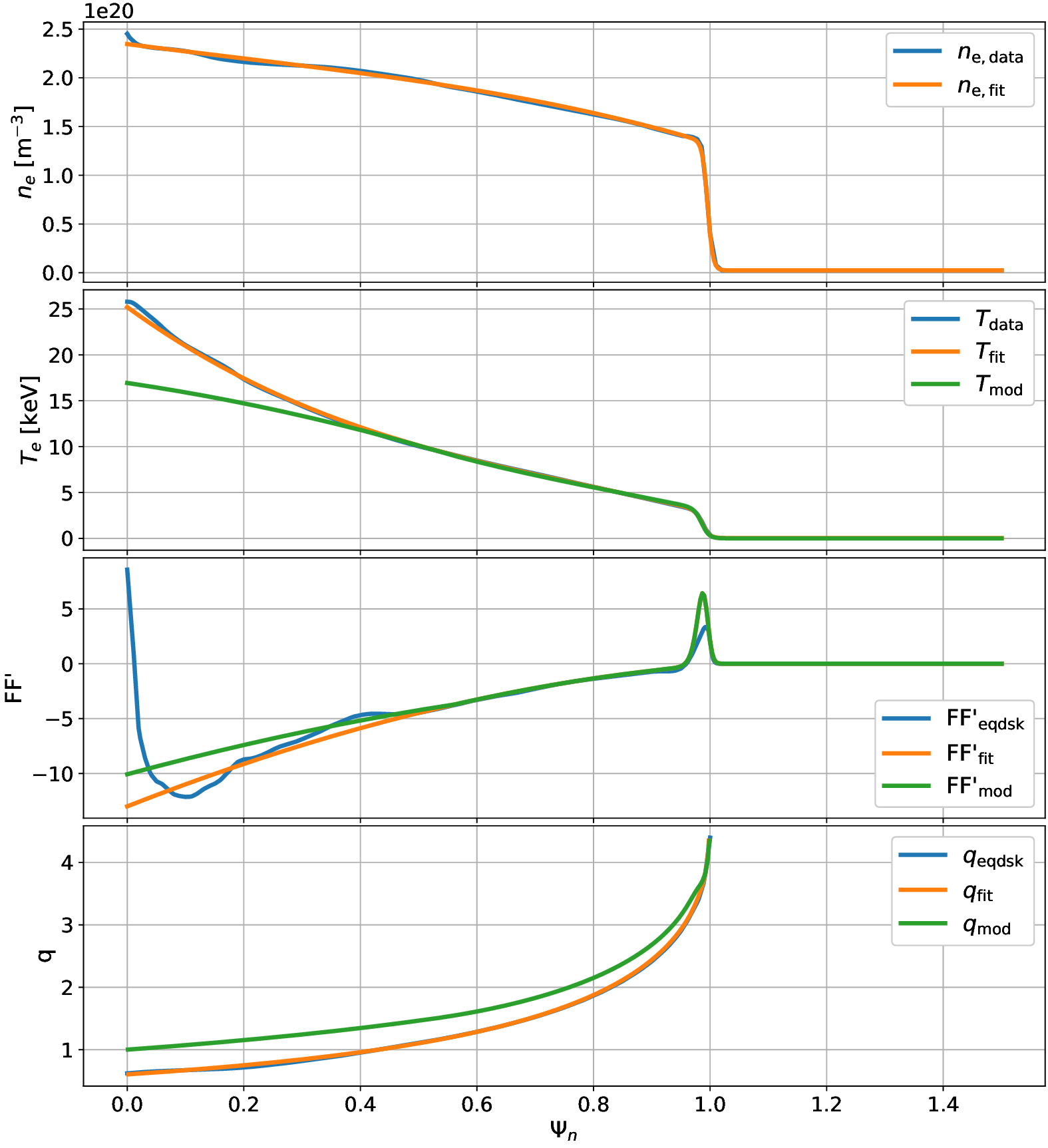}
    \caption{Comparison of the equilibrium profiles for electron density $n_e$, electron temperature $T_e$, FF', and safety factor $q$ (from top to bottom) as a function of the normalized poloidal flux $\psi_n$. The plots display the CREATE-NL equilibrium data (blue) and their fitted approximations (orange). The fitted $n_e$ profile was not modified, while for $T_e$, FF', and $q$, the modified MHD-stable profiles employed in the simulations are shown in green.}
   \label{fig:mod_eq}
\end{figure}

\begin{figure*}[t!]
    \centering
    \includegraphics[width=0.75\textwidth]{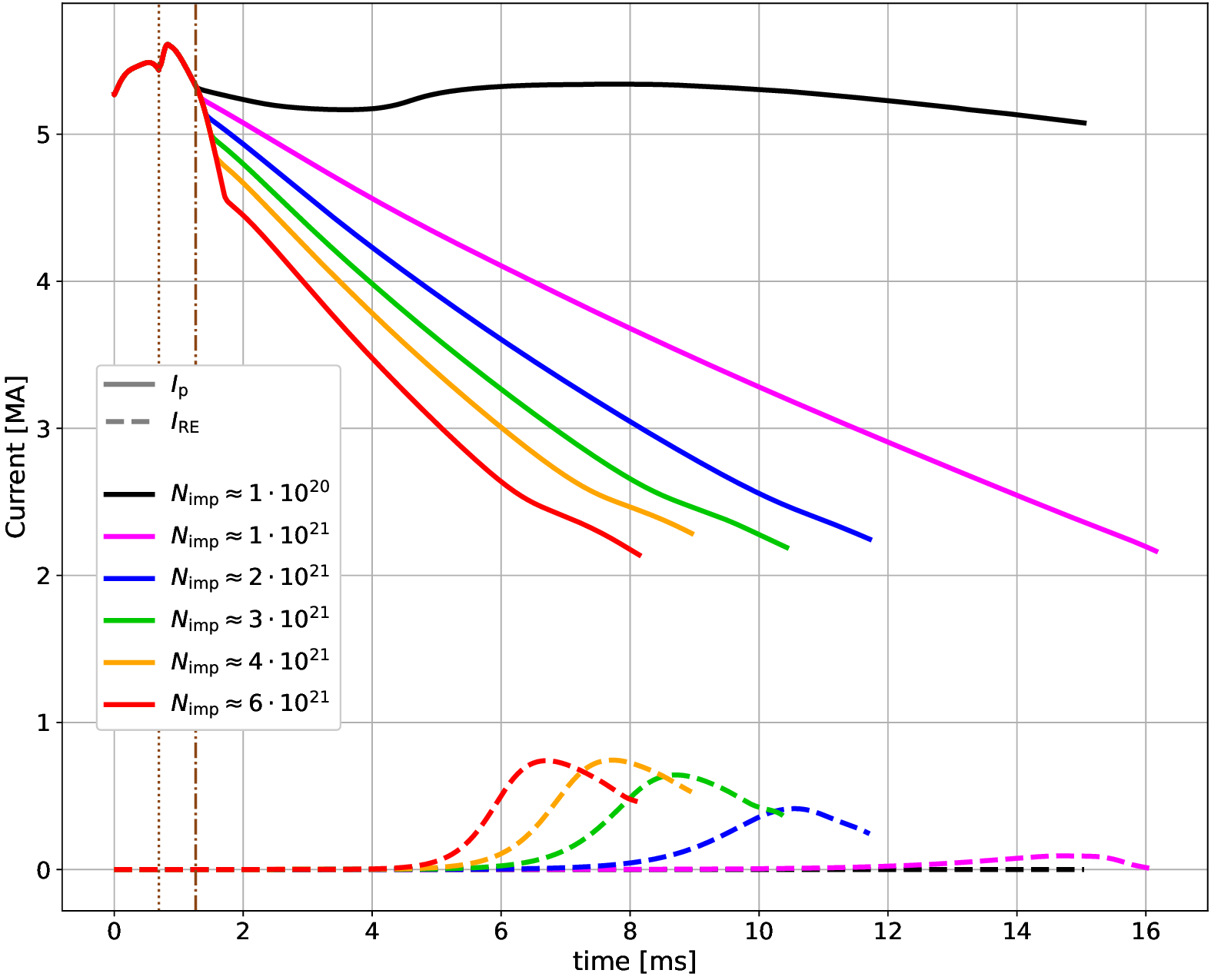}
    \caption{Time evolution of the total plasma current $I_\text{p}$ (solid lines) and RE current $I_\text{RE}$ (dashed lines) for a small initial RE seed of $I_\text{seed} = 5.5$~A. The colors correspond to varying levels of injected neon impurities ($N_\text{imp}$), ranging from $1\cdot10^{20}$ to $6\cdot10^{21}$ particles. Thin brown vertical lines indicate the ATQ phases: the dotted line separates the thermal collapse from the current flattening phase, while the dashed-dotted line marks the onset of impurity injection. The transition to the self-consistent CQ is identified by the change in slope of the plasma current traces following the dashed-dotted line. Since impurities are injected at a fixed rate, higher $N_\text{imp}$ values require longer injection times, delaying the CQ onset. Despite the very small initial seed, macroscopic RE beams reaching $\approx 0.7$ MA form at high impurity levels ($N_\text{imp} \gtrsim 3\cdot10^{21}$), demonstrating the high avalanche gain of the full power scenario of DTT.}
    \label{fig:seed_6}
\end{figure*}

In this work we focus our investigation on the DTT single null full power configuration (scenario E1 \cite{Casiraghi2023}), characterized by a nominal plasma current of $I_p = 5.5$ MA, a toroidal magnetic field of $B_\phi = 5.85$ T, and an auxiliary heating power of approximately 45 MW. Standard integrated modeling of this scenario typically predicts a safety factor $q < 1$ for a large portion of the core, possibly causing large sawteeth crashes \cite{Romanelli2024}. While work on sawtooth control optimization is ongoing, we focus here on an MHD-stable magnetic equilibrium ($q > 1$), obtained by modifying the FF' and temperature profiles of a CREATE-NL \cite{Albanese2015} equilibrium for the full power scenario. Figure \ref{fig:mod_eq} displays the CREATE-NL equilibrium data (blue) and their fitted approximations (orange) for the electron density $n_e$, electron temperature $T_e$, FF', and safety factor $q$ (from top to bottom) as a function of the normalized poloidal flux $\psi_n$. The fitted $n_e$ profile was not modified, while for $T_e$, FF', and $q$, the modified MHD-stable profiles employed in the simulations are also shown (green). These profile modifications are justified by the specific nature of the ATQ procedure. The initial thermal collapse, which reduces the temperature by several orders of magnitude, renders the exact shape of the initial profile largely unimportant. Furthermore, the subsequent current-flattening phase inevitably reshapes the $q$ profile in the core region, overriding the initial equilibrium adjustments. As a result, the changes made to ensure initial MHD stability do not significantly impact the CQ or RE avalanche dynamics. Accordingly, the equilibrium adopted for all simulations combines the modified profiles for $T_e$, FF', and $q$ (green lines) with the fitted electron density (orange line) shown in Figure \ref{fig:mod_eq}. Future work will ideally broaden this investigation by considering alternative, intrinsically MHD-stable DTT equilibria, provided such configurations become available, to assess the sensitivity of the RE avalanche to the initial magnetic configuration.
Since accurately estimating the amplitude of primary RE generation mechanisms is challenging \cite{Breizman2019}, often yielding values that span several orders of magnitude, we treat the initial RE seed current as a scanning parameter. To provide a comprehensive RE safety assessment for the DTT full power scenario, we select representative seed currents covering a wide parameter space, setting $I_\text{seed}$ to $10^{-6}$, $10^{-4}$, and $10^{-2}$ of the initial plasma current $I_p = 5.5$ MA (yielding 5.5 A, 550 A, and 55 kA, respectively). Additionally, we explicitly include a seed value of 20 kA, identical to the maximum seed used in our previous Day-0 assessment \cite{Emanuelli2025}, to facilitate a direct comparison between the two operating points, allowing us to isolate the effects of the higher current regime. We emphasize that the upper bound of 55 kA constitutes a substantial initial population of REs. Were the seed an order of magnitude larger, it would pose a severe threat even with minimal subsequent avalanche multiplication. Such a scenario would shift the focus from CQ dynamics to primary RE generation physics, which lies outside the scope of this work.
Parallel to the seed scan, we vary the amount of neon impurities injected, which critically influences the CQ dynamics and RE avalanche gain. As highlighted in \cite{Hesslow2019}, bound electrons in partially ionized impurities can act as efficient targets for knock-on collisions, potentially enhancing the avalanche growth rate significantly. Furthermore, the impurity content dictates the radiative cooling rate, which critically determines the plasma resistivity and the subsequent induced electric field that accelerate electrons. In this study, the injected impurity inventory is varied from $1 \cdot 10^{20}$ to $6 \cdot 10^{21}$ particles (distributed uniformly within a computational volume of $\approx 52$ $\text{m}^3$). These values are consistent with the injection of a 12 mm pellet with a 20\% neon fraction (containing roughly $10^{23}$ neon particles), considering an assimilation of less than 10\% before the TQ. In contrast to RE seeds, the impurity content is an experimentally controllable parameter, directly determined by the quantity of material injected by the DMS. The primary mitigation strategy for DTT relies on shattered pellet injection (SPI), while massive gas injection (MGI) could be considered before full commissioning of the SPI system. By systematically exploring the parameter space defined by these seed currents and impurity levels, we aim to provide a comprehensive safety assessment regarding RE beam formation in DTT. This broad scan allows us to characterize the dynamics of RE avalanche across widely different regimes. Future studies may extend this analysis to an even broader parameter space, consider different impurity species, or implement more realistic spatial distributions of impurities within the computational domain.

\section{Results and Discussion}
\label{sec:results}

This section presents the results of the JOREK simulations for the DTT full power scenario ($I_p = 5.5$ MA). The analysis focuses on the sensitivity of RE beam formation to the initial pre-existing RE seed population ($I_\text{seed}$) and the amount of impurities injected by the DMS ($N_\text{imp}$). Following the ATQ procedure described in Section \ref{sec:model}, the plasma enters the CQ phase. Here, the electron temperature $T_e$ evolves self-consistently inside JOREK, balancing Ohmic heating against impurity radiation. This balance determines the plasma resistivity and, consequently, the strength of the parallel electric field $E_{\parallel}$ that drives the RE avalanche.

We begin by considering the case with the lowest initial RE seed current (5.5 A, representing a fraction $I_\text{seed}/I_p = 10^{-6}$). This scenario mimics conditions where primary RE generation mechanisms are weak or have been effectively suppressed prior to the onset of the avalanche phase. The time evolution of the plasma current ($I_p$, solid lines) and the generated RE current ($I_\text{RE}$, dashed lines) is shown in Figure \ref{fig:seed_6}. A striking feature of the full power scenario is its susceptibility to macroscopic RE beam formation even from such tiny RE seeds. For low impurity contents ($N_\text{imp} \approx 1\cdot10^{20}$, black curves), the radiative losses are insufficient to maintain a low electron temperature. Consequently, the plasma resistivity remains moderate, resulting in a current decay that is extremely slow. In this case, the induced electric field remains below the critical threshold required to overcome the collisional drag and sustain a significant multiplication of REs. As a result, the RE current (dashed black curve) remains negligible.
\begin{figure}[t!]
    \centering
    \includegraphics[width=0.95\columnwidth]{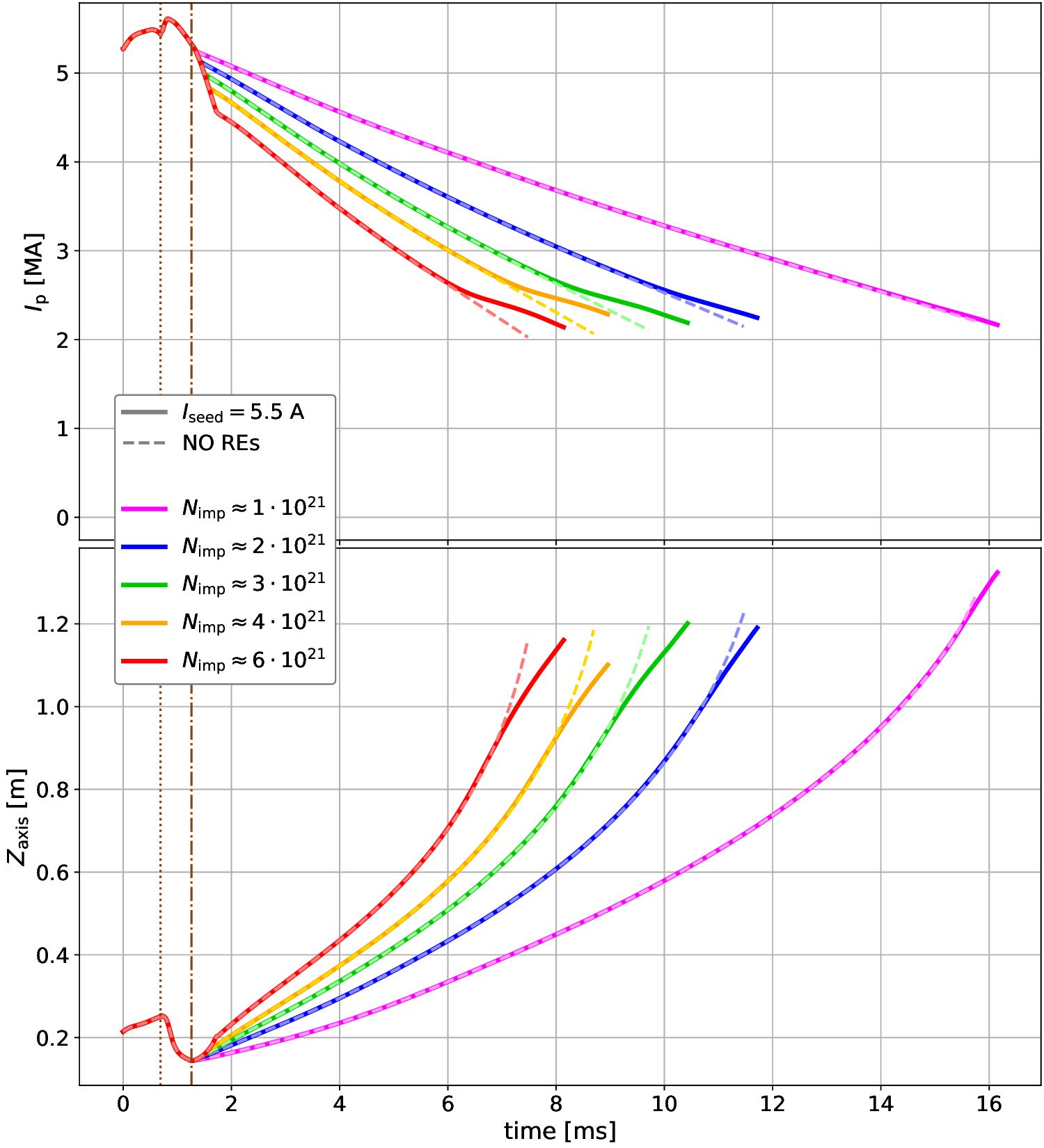}
    \caption{Comparison of the total plasma current $I_p$ (top) and the vertical position of the magnetic axis $Z_\text{axis}$ (bottom) for the smallest-seed case ($I_\text{seed}=5.5$ A, solid thick lines) and reference simulations without REs (dashed thin lines, obtained by numerically suppressing the avalanche source term). The colors correspond to varying levels of $N_\text{imp}$, and thin brown vertical lines demarcate the ATQ phases, consistent with Figure \ref{fig:seed_6}. The bottom panel reveals that higher impurity levels lead to steeper vertical drifts. For this small initial seed, the RE beam takes several milliseconds to become macroscopic and deviate from the RE-free dynamics, eventually slowing the vertical displacement.}    
    \label{fig:z_seed6}
\end{figure}
\begin{figure}[t!]
    \centering
    \includegraphics[width=0.95\columnwidth]{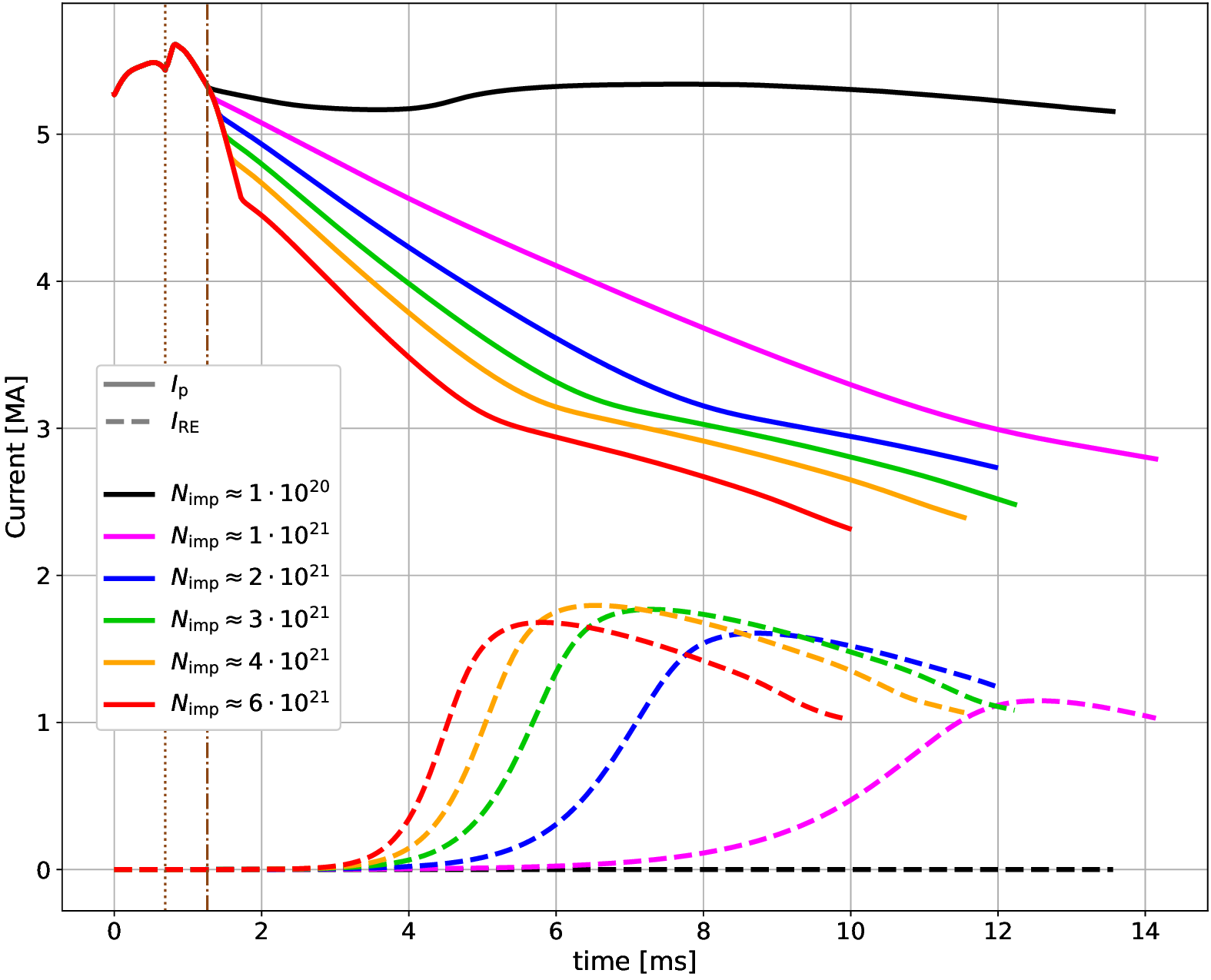}
    \caption{Time evolution of the total plasma current $I_p$ (solid lines) and RE current $I_\text{RE}$ (dashed lines) for an intermediate initial RE seed of $I_\text{seed} = 550$ A. The colors correspond to varying levels of $N_\text{imp}$, and thin brown vertical lines demarcate the ATQ phases, consistent with Figure \ref{fig:seed_6}. Significant RE beam formation occurs at lower impurity thresholds compared to the 5.5 A case. The highest RE current reaches approximately $1.8$ MA (accounting for almost $60\%$ of the total current).}    
    \label{fig:seed_4}
\end{figure}
\begin{figure}[t!]
    \centering
    \includegraphics[width=0.95\columnwidth]{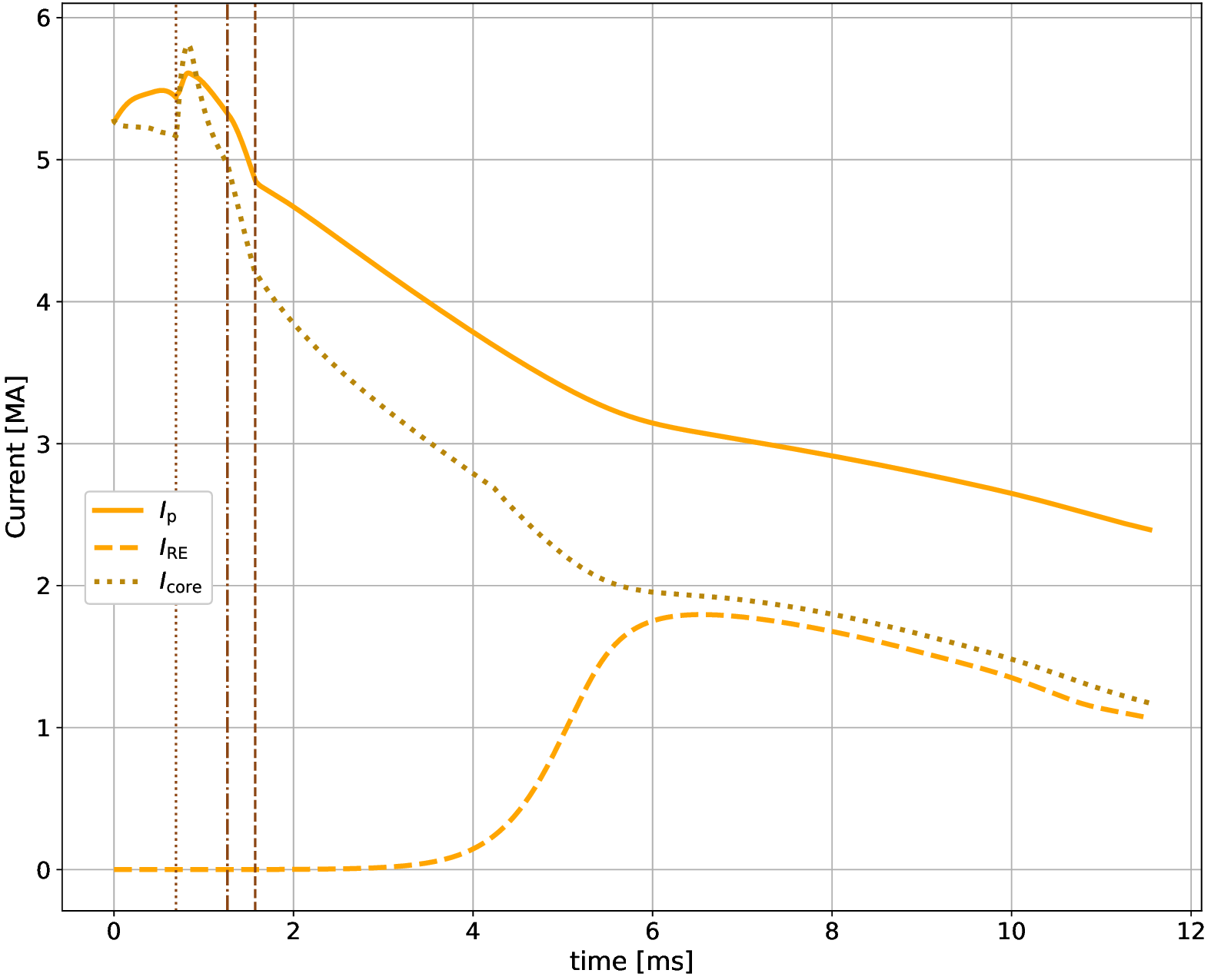}
    \caption{Current decomposition for the intermediate seed scenario ($I_\text{seed} = 550$ A) with $N_\text{imp} \approx 4\cdot10^{21}$ particles. The solid ($I_p$) and dashed ($I_\text{RE}$) lines correspond to the orange traces in Figure \ref{fig:seed_4}, while the gold dotted line shows the total current in the confined region ($I_\text{core}$). Thin brown vertical lines indicate the ATQ phases as in previous figures, with the addition of a dashed vertical line at $t \approx 1.6$ ms marking the end of the impurity injection and the start of the self-consistent CQ for this specific scenario. The proximity of the $I_\text{core}$ and $I_\text{RE}$ traces after $t \approx 6$ ms indicates that the confined current is dominated by the RE population, implying that the thermal current ($I_p - I_\text{RE}$) corresponds largely to halo currents.}
    \label{fig:4e21_current}
\end{figure}
However, as the impurity content is increased, the CQ dynamics changes radically. For $N_\text{imp} \gtrsim 1\cdot10^{21}$ (the colored curves), the enhanced radiation clamps the post-TQ temperature to lower and lower values (typically $\approx 5-15$ eV). This drastic increase in plasma electrical resistivity leads to faster CQs. As the current decay rate increases, the induced toroidal electric field intensifies, leading to a more vigorous avalanche multiplication that causes the RE current to peak earlier in the simulation. Moreover, the resulting $E_\parallel$ in these cases is sufficient to trigger massive avalanche amplification of REs. Specifically, in the cases with the highest impurity contents ($3\cdot10^{21} \lesssim N_\text{imp} \lesssim 6\cdot10^{21}$), the 5.5 A seed is amplified by more than five orders of magnitude, resulting in a macroscopic RE beam of $I_\text{RE} \approx 0.7$ MA. Although substantial, this RE current remains limited to a fraction of the total plasma current, as the rapid vertical displacement of the plasma (bottom panel in Figure \ref{fig:z_seed6}) leads to the scrape-off of the RE population against the first wall, accelerating the reduction of the confined plasma volume and limiting further growth of the RE population. Comparing the solid lines ($I_\text{seed}=5.5$ A) with the dashed lines (no REs) in Figure \ref{fig:z_seed6}, we observe that the curves diverge only in the final milliseconds of the simulations, indicating that the RE beam must grow to macroscopic levels before effectively slowing down the vertical displacement. An impurity saturation effect appears for $N_\text{imp} \approx 6\cdot10^{21}$ (red curve), where the RE current peak diminishes despite occurring earlier (Figure \ref{fig:seed_6}). This confirms that at such impurity levels the enhanced RE avalanche production is counterbalanced by the macroscopic losses driven by the accelerated vertical displacement (see Figure \ref{fig:z_seed6}). This result contrasts sharply with the Day-0 simulations \cite{Emanuelli2025}, where the lowest seeds invariably resulted in safe scenarios, with a maximum amplification (for the case with $N_\text{imp} \approx 3\cdot10^{21}$) of $G_\text{av}\approx500$. At $I_p=5.5$ MA, though, the avalanche gain $G_\text{av}$ is so large ($\approx1.3\cdot10^{5}$) that having a very small RE seed is not guaranteed to be a safe condition if the disruption dynamics creates a sufficiently strong electric field. 

\begin{figure}[b!]
    \centering
    \includegraphics[width=0.95\columnwidth]{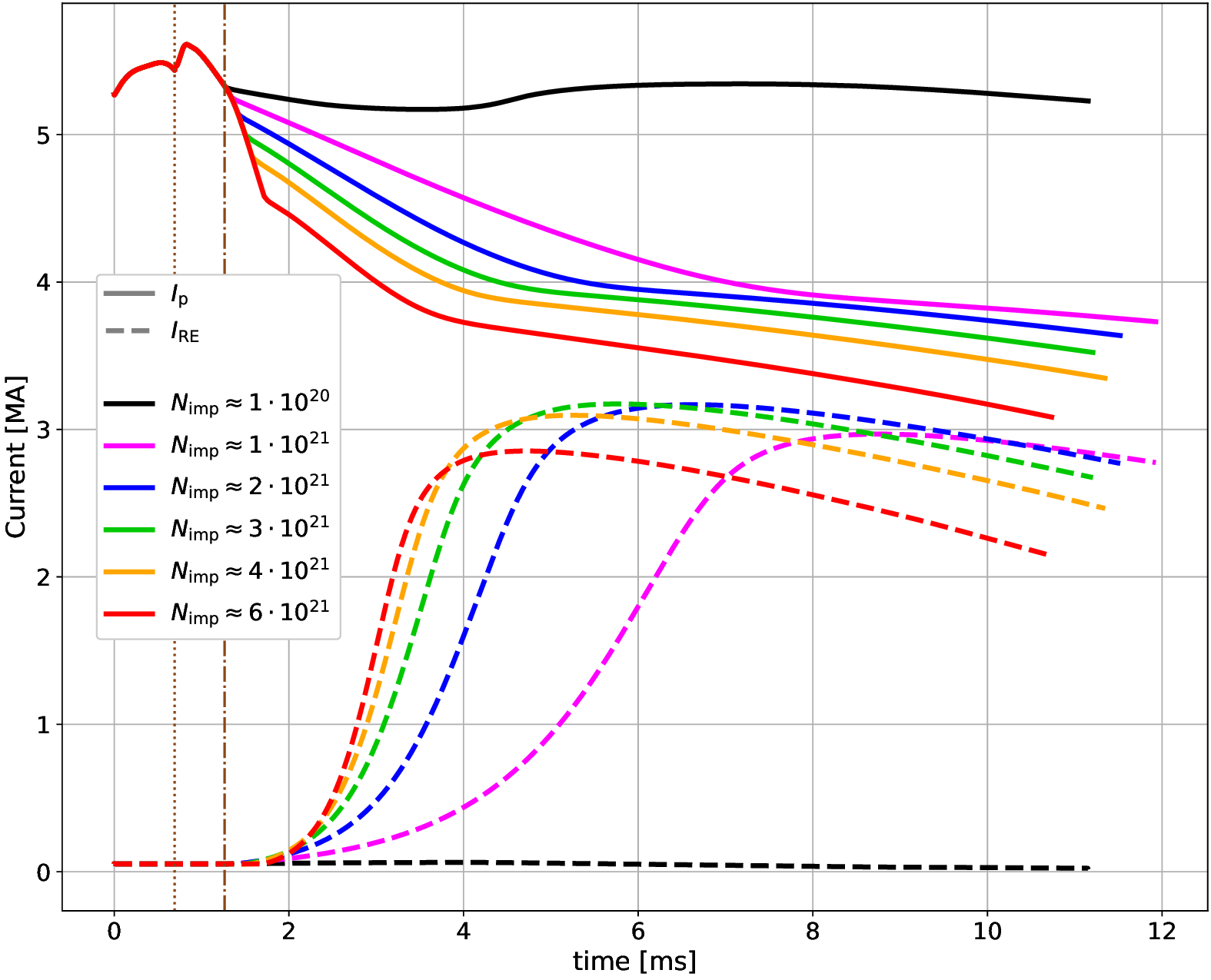}
    \caption{Time evolution of the total plasma current $I_p$ (solid lines) and RE current $I_\text{RE}$ (dashed lines) for a large initial RE seed of $I_\text{seed} = 55$ kA. The colors correspond to varying levels of $N_\text{imp}$, and thin brown vertical lines demarcate the ATQ phases, consistent with Figure \ref{fig:seed_6}. Within a few milliseconds, the RE current evolves into $\approx 80\%$ of the total current, even for impurity levels as low as $N_\text{imp} \approx 1\cdot10^{21}$. The maximum RE current saturates at $\approx 3.2$ MA, significantly slowing down the CQ.}
    \label{fig:seed_2}
\end{figure}
\begin{figure}[t!]
    \centering
    \includegraphics[width=0.95\columnwidth]{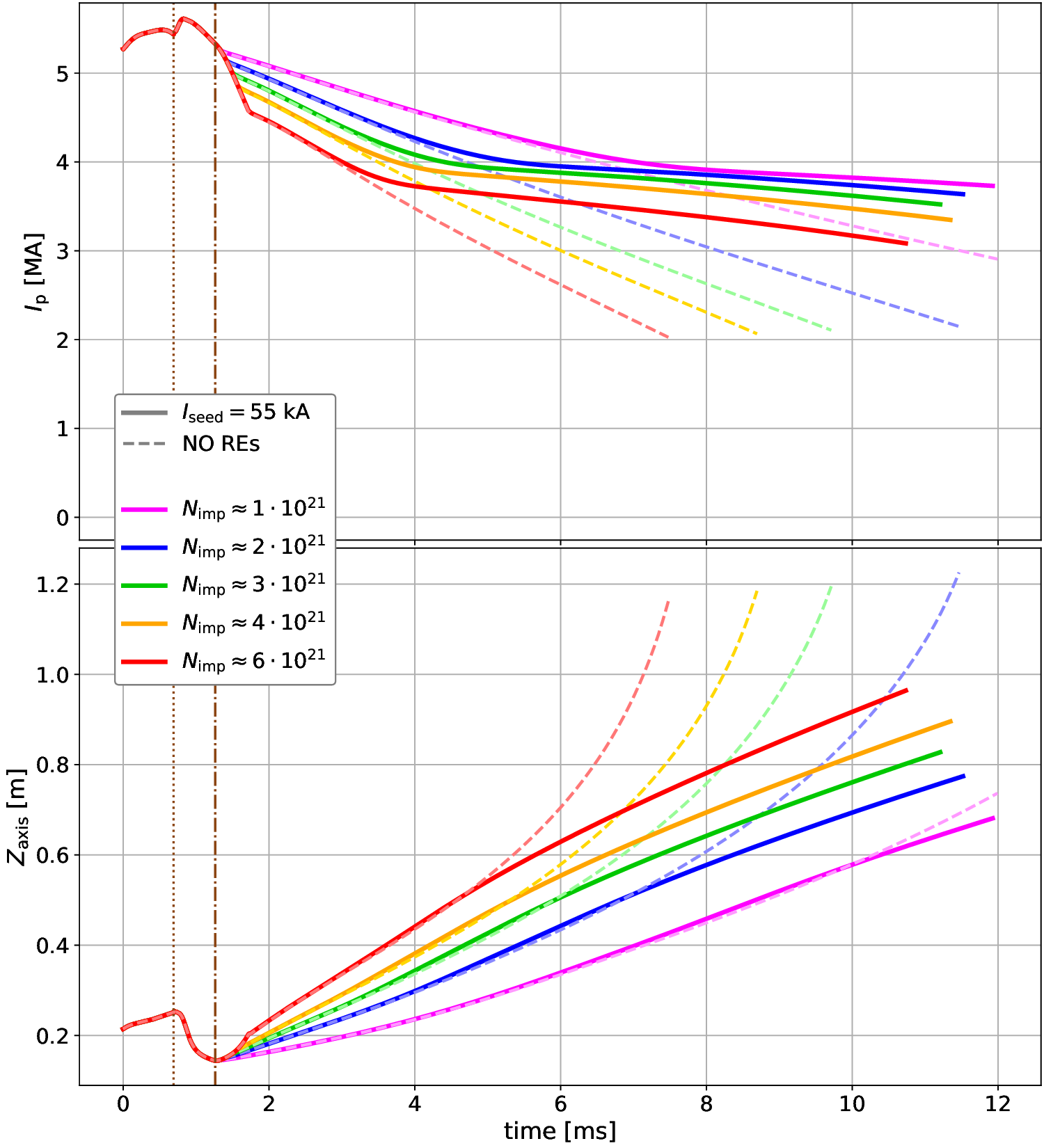}
    \caption{Comparison of the total plasma current $I_p$ (top) and the vertical position of the magnetic axis $Z_\text{axis}$ (bottom) for the largest-seed case ($I_\text{seed}=55$ kA, solid thick lines) and reference simulations without REs (dashed thin lines, obtained by numerically suppressing the avalanche source term). The colors correspond to varying levels of $N_\text{imp}$, and thin brown vertical lines demarcate the ATQ phases, consistent with Figure \ref{fig:seed_2}. The high current carried by the REs in these scenarios causes a substantial reduction in the velocity of the vertical displacement and of the CQ compared to the RE-free scenarios.}
    \label{fig:z_seed2}
\end{figure}

\begin{figure*}[t!]
    \centering
    \includegraphics[width=0.9\textwidth]{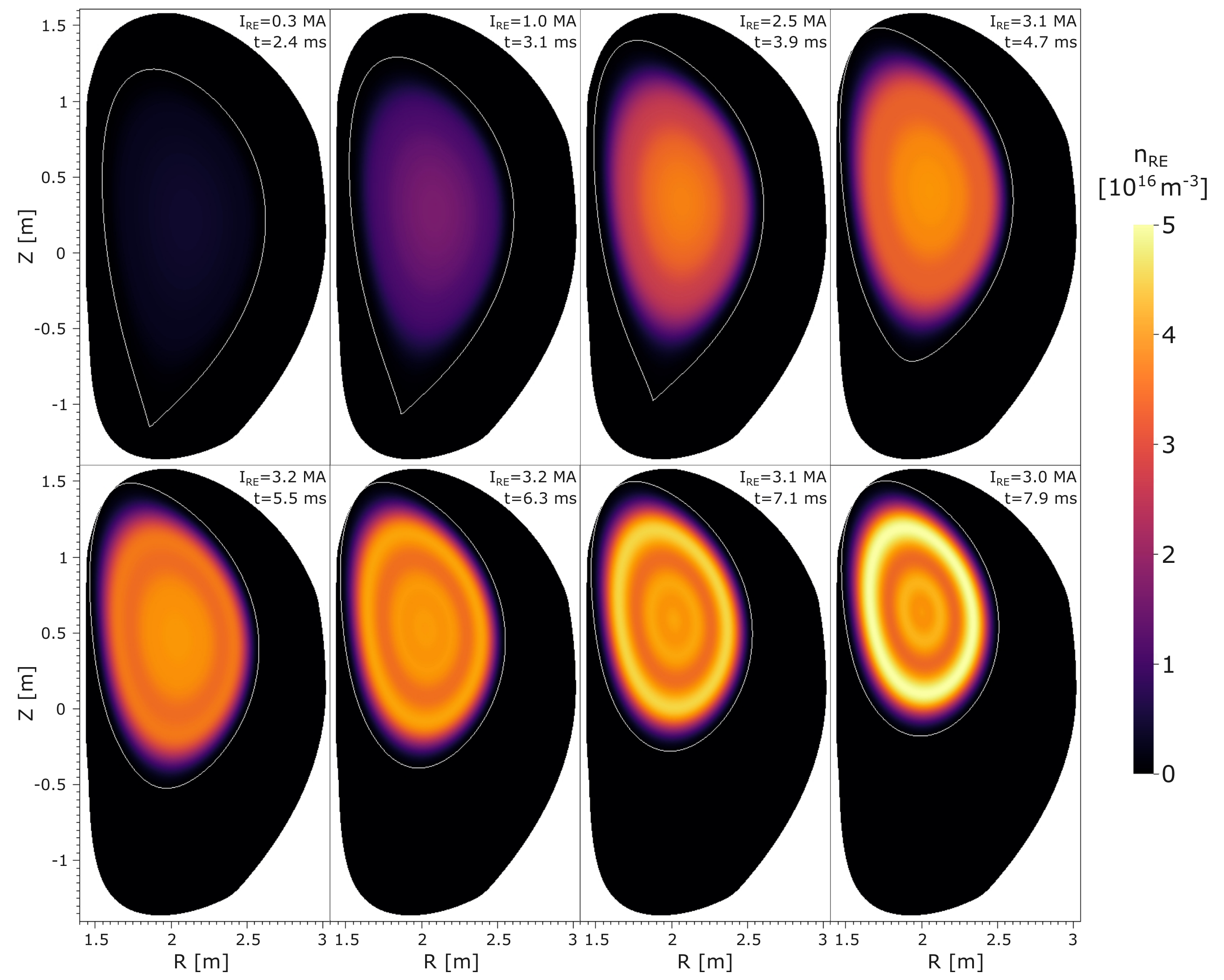}
    \caption{Poloidal cross-section evolution of the RE density $n_\text{RE}$ (in $10^{16}$\thinspace m$^{-3}$) for the large-seed case ($I_\text{seed}=55$ kA) with $N_\text{imp} \approx 3 \cdot 10^{21}$ (corresponding to the green curves in Figure \ref{fig:seed_2} and the green solid lines in Figure \ref{fig:z_seed2}). The white contour is the last closed flux surface, separating the confined region from the open field lines. As the vertical displacement progresses, the plasma volume is scraped off by the first wall. Although the core RE density increases due to the avalanche, the total RE current ($I_\text{RE}$, top right of each panel) eventually saturates (reaching its maximum at $t \approx 5.8$ ms) and decays due to the reduction of the confined volume.}    
    \label{fig:RE_evolution}
\end{figure*}

Figure \ref{fig:seed_4} illustrates the current evolution for an intermediate RE seed of 550 A ($I_\text{seed}/I_p = 10^{-4}$), serving as a proxy for scenarios with moderate primary generation. The qualitative dependence on the impurity content remains consistent with the low-seed case: lower impurity levels ($N_\text{imp} \approx 1\cdot10^{20}$, black lines) result in negligible RE generation, as the electric field is insufficient to drive the RE avalanche.
However, owing to an initial seed two orders of magnitude larger than in the previous case, the RE populations reach their peak values earlier and the resulting beam currents are substantially higher. This difference is particularly evident at intermediate impurity levels ($1\cdot10^{21} \lesssim N_\text{imp} \lesssim 2\cdot10^{21}$, pink and blue curves). In particular, with $N_\text{imp} \approx 4\cdot10^{21}$, the RE current reaches approximately $1.8$ MA, accounting for almost $60\%$ of the total current. To better characterize the current distribution for this impurity case, Figure \ref{fig:4e21_current} compares $I_p$ and $I_\text{RE}$ from Figure \ref{fig:seed_4} (orange lines) with the confined current ($I_\text{core}$) extracted from the same simulation. As the avalanche proceeds, $I_\text{RE}$ approaches $I_\text{core}$ closely, indicating that the core current is predominantly carried by REs. This implies that the thermal current (visible as the difference between $I_p$ and $I_\text{RE}$) corresponds largely to halo currents flowing on open field lines, where the electron temperature is typically found to be $\approx 5$ eV in the simulations. At this stage, the CQ dynamics is dominated by the RE beam, which significantly slows the current decay (an effect only marginally observable with $I_\text{seed}= 5.5$ A). This $I_p$ decay is driven primarily by the vertical displacement of the plasma column rather than by resistive dissipation. As the plasma drifts vertically, the outer flux surfaces are successively scraped off by the first wall, leading to a gradual loss of the RE population (visualized in detail in Figure \ref{fig:RE_evolution}). This interaction poses a severe threat to the PFCs, as the substantial energy stored in the RE beam is deposited onto localized wall regions. Although studying an MHD-induced broadening of the deposition area might show paths towards a benign RE termination scenario, it lies outside the scope of this work.
Ultimately, the accelerated vertical displacement at higher impurity levels limits the beam growth: for $N_\text{imp}$ exceeding $4\cdot10^{21}$ (red lines), the maximum RE current begins to decrease, consistent with the saturation effect observed in the low-seed analysis.

Figure \ref{fig:seed_2} presents the results for the largest initial seed ($I_\text{seed} = 55$ kA, $I_\text{seed}/I_p = 10^{-2}$), representing the upper bound of our investigation. This corresponds to a scenario where primary generation mechanisms have converted a significant fraction of the plasma current into REs immediately after the TQ. As noted in Section \ref{sec:model}, increasing the seed magnitude beyond this limit would pose a serious threat even with minimal multiplication, thus shifting the dominant physical mechanism from avalanche multiplication to primary generation. Although the qualitative behavior mirrors the trends observed for the intermediate seed, the drastic increase in $I_\text{seed}$ significantly reduces the time to reach RE current saturation, while also yielding a substantially higher peak $I_\text{RE}$. In fact, there is almost no lag time between the start of the CQ and the formation of a macroscopic RE beam, which drastically slows the CQ. The impact of this high RE current on vertical stability is evident in Figure \ref{fig:z_seed2}: the RE beam significantly slows the vertical displacement compared to the RE-free reference cases (dashed lines), where the vertical position ($Z_\text{axis}$) rises much faster.
Within a few milliseconds, the RE current rapidly constitutes $\approx 80\%$ of the total current, even for impurity levels as low as $N_\text{imp} \approx 1\cdot10^{21}$. The largest RE current for this 55 kA seed is now obtained with $N_\text{imp} \approx 3\cdot10^{21}$ (green curves), reaching $I_\text{RE} \approx 3.2$ MA (nearly 60\% of the pre-disruption plasma current) at $t \approx 5.8$ ms. For higher levels of impurities (orange and red lines), the maximum RE current starts to decrease. This lower impurity threshold with respect to the previous cases confirms that the wall-scraping mechanism of the RE beam is accentuated in this scenario: the confined plasma current is almost entirely converted to RE current, and the slow current decay is primarily driven by the vertical drift of the plasma column. Figure \ref{fig:RE_evolution} illustrates the process for the case with $N_\text{imp} \approx 3\cdot10^{21}$: as the vertical displacement proceeds, the confined plasma volume (delimited by the last closed flux surface, white contour) shrinks significantly. Consequently, even though the RE density in the core continues to rise due to the avalanche, the total RE current ($I_\text{RE}$) saturates and eventually decreases as the flux surfaces are scraped off by the first wall. 
It is worth noting that even in this extreme $I_\text{seed}$ case, the simulation with the lowest impurity content ($N_\text{imp} \approx 1\cdot10^{20}$, black lines) remains RE-free. This stresses the critical role of the impurities: without sufficient cooling to raise the background resistivity, even a massive 55 kA seed will eventually decay due to collisional drag and radial diffusion, rather than avalanching. Nevertheless, the thermal loads in these virtually unmitigated scenarios would likely be unacceptable.

\begin{figure}[t!]
    \centering
    \includegraphics[width=0.95\columnwidth]{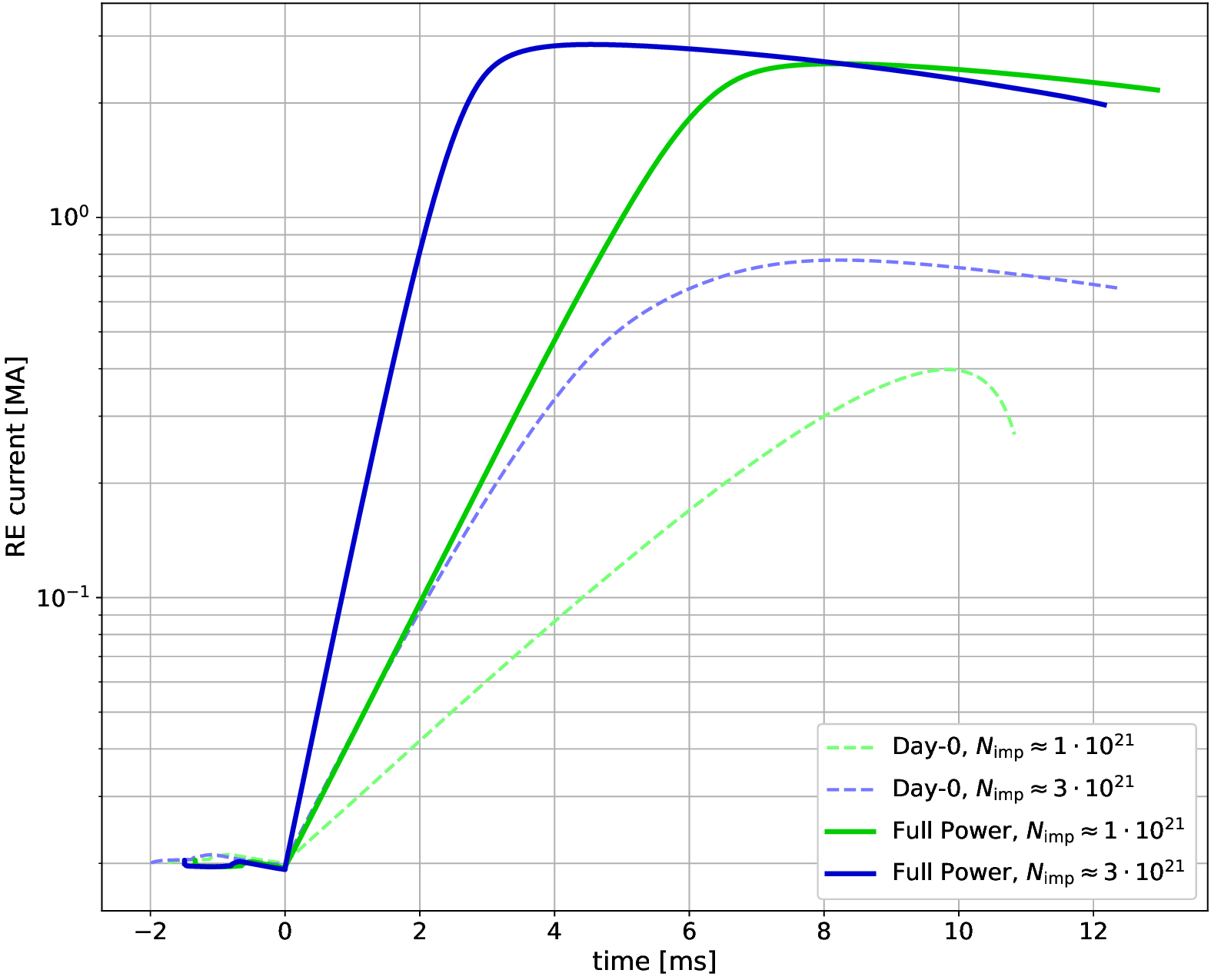}
    \caption{Direct comparison of the RE current evolution in the Day-0 scenario ($I_p \approx 2$ MA, dashed lines, data adapted from \cite{Emanuelli2025}) and the full power scenario ($I_p \approx 5.5$ MA, solid lines) for a fixed initial seed of 20 kA, with $t=0$ ms aligned to the start of the CQ. Green and blue lines correspond to $N_\text{imp} \approx 1\cdot10^{21}$ and $3\cdot10^{21}$, respectively. The logarithmic scale highlights that the full power scenario is characterized by much steeper slopes and higher peak currents, reflecting the exponential disparity in avalanche amplification.}
    \label{fig:day0_FP}
\end{figure}

To isolate the impact of the operational scenario on the generation of RE we performed a direct comparison between Day-0 ($I_p=2$ MA) and full power ($I_p=5.5$ MA), visualized in Figure \ref{fig:day0_FP}. The comparison relies on a 20 kA RE seed and on impurity levels equal to $N_\text{imp} \approx 1\cdot10^{21}$ and $3\cdot10^{21}$. These parameters were selected to compare the worst-case scenarios for both regimes, being the only cases found to yield a significant RE beam in the Day-0 scenario. The Day-0 simulation data used for this comparison were previously reported in \cite{Emanuelli2025}.
The logarithmic scale in Figure \ref{fig:day0_FP} reveals the profound difference in RE generation of the two scenarios. The dashed lines, representing the Day-0 case, show that $I_\text{RE}$ increases only marginally from the 20 kA seed, even for high impurity levels, always remaining below 1 MA and, hence, posing minimal risks to the PFCs of DTT \cite{Emanuelli2025}. In stark contrast, the full power simulations (solid lines) exhibit explosive growth, with curve slopes that are much steeper than the Day-0 counterparts and reaching higher peak RE currents (nearly 3 MA). This confirms the risks highlighted by the avalanche gain analysis of the 5.5 A seed: in the presence of a relatively high RE seed, the very large amplification potential of the full power scenario ($G_\text{av}\approx1.3\cdot10^{5}$) manifests itself as a rapid and robust conversion of the core plasma current into RE current.

The results of this parametric scan delineate a significantly narrower ``RE safe'' operational window for the DTT full power scenario compared to Day-0. Although the safety against RE beams of the 2 MA scenario was largely insensitive to the specific mix of initial RE seeds and impurities (since almost all reasonable paths led to negligible RE currents), at 5.5 MA the system is highly sensitive to the combination of these parameters. 
We can identify three distinct regimes from the parametric scan (Figures \ref{fig:seed_6}, \ref{fig:seed_4}, and \ref{fig:seed_2}).
\begin{enumerate}
    \item Benign resistive decay ($N_\text{imp} \lesssim 1\cdot10^{20}$). In this regime, the quantity of impurities injected is too low to extensively cool the plasma. The plasma resistivity remains relatively low, preventing the formation of the large electric fields required for avalanching. While effective at suppressing REs, the capacity of this regime to radiate sufficiently and mitigate heat loads or halo current forces requires further analysis, notwithstanding the mechanical robustness of the DTT vacuum vessel against 750 worst-case disruptions \cite{Romanelli2024}.
    \item Avalanche transition ($1\cdot10^{20} < N_\text{imp} \lesssim 1\cdot10^{21}$). This regime marks the onset of the avalanche mechanism, characterized by positive but moderate amplification. A low RE seed (5.5 A) can initiate an avalanche (Figure \ref{fig:seed_6}, pink line), but the gain is insufficient to produce a macroscopic RE beam before the plasma column is scraped off against PFCs. Conversely, larger seeds can compensate for the moderate gain, resulting in significant RE currents ((pink lines in Figures \ref{fig:seed_4} and \ref{fig:seed_2}). In this regime, the safety against RE beams depends on the magnitude of the primary seed, whose accurate prediction remains a significant challenge.
    \item Macroscopic beam formation ($N_\text{imp} \gtrsim 2\cdot10^{21}$). In this high-impurity regime, the safety margin effectively vanishes. The avalanche gain is substantial enough to elevate even microscopic seeds to macroscopic RE currents. This result suggests that massive impurity injection for thermal mitigation carries a severe penalty in terms of RE risk for the full power scenario, irrespective of the initial RE seed current.
\end{enumerate}
It must be noted that these simulations, being 2D (toroidally symmetric), represent a ``worst-case'' upper bound regarding RE avalanche. In a 3D disruption, MHD instabilities may destroy the magnetic flux surfaces, creating stochastic field lines that enhance RE losses through radial transport. Although 3D simulations are required to quantify this loss, the fact that 2D simulations predict beams of up to $\approx 3.2$ MA indicates that the source term is extremely strong and relying solely on stochastic losses to suppress such a vigorous avalanche may not be sufficient.

\section{Conclusions}
\label{sec:conclusion}
This work has extended the safety analysis of RE avalanches in DTT to the full power scenario ($I_{p}=5.5$ MA). Building upon previous assessments of the Day-0 commissioning phase ($I_{p}=2$ MA), we employed the non-linear MHD code JOREK and the resistive wall code STARWALL to simulate the current quench dynamics under conditions relevant to reactor-scale operations. By performing an extensive parametric scan of initial RE seed currents (ranging from 5.5 A to 55 kA) and injected neon impurity levels, we have quantified the risk of RE beam formation in this high-current scenario of DTT.
The comparative analysis reveals a fundamental divergence from the Day-0 results. While the 2 MA scenario was characterized by a broad safety margin in which RE generation was typically negligible, the 5.5 MA scenario exhibits a high sensitivity to the input parameters. Specifically, the exponential increase in the avalanche gain allows even minute seed populations ($I_\text{seed}/I_{p} = 10^{-6}$) to be amplified into macroscopic beams if the impurity density is sufficiently high to drive a fast current quench ($N_\text{imp} \gtrsim 2 \cdot 10^{21}$).

Our analysis identifies three distinct physical regimes for the full power scenario. For low impurity levels ($N_\text{imp} \lesssim 1\cdot10^{20}$), the induced electric field is too small to sustain a RE avalanche; however, the reduced radiative cooling may result in thermal loads and electromagnetic forces that the PFCs and supporting structures cannot withstand. For higher values of $N_\text{imp}$ up to $\approx 1\cdot10^{21}$, the avalanche multiplication becomes increasingly dominant. In this transition zone, the safety of the device against REs is highly dependent on the magnitude of the initial RE seed, which originates from primary RE generation mechanisms (e.g., Dreicer or hot-tail). Forecasting safety in this regime is challenging due to the inherent difficulty in predicting the primary seed amplitude. Finally, for $N_\text{imp} \gtrsim 2\cdot10^{21}$, the safety margin is drastically reduced: even very small initial seeds are amplified into macroscopic RE beams, while larger seeds facilitate the conversion of almost all the confined plasma current into RE current. In such cases, the CQ is effectively stalled by the highly conductive RE population. Consequently, the discharge termination is governed by the vertical displacement of the plasma column and its subsequent interaction with the first wall, where the substantial energy carried by the RE beam is deposited onto localized surfaces, representing a severe threat to the integrity of the PFCs.

Given the severe risks identified in the high-current scenario of DTT, future studies will focus on identifying an optimal disruption mitigation strategy. While this study considered pure neon injection to isolate the impurity dependence, a practical solution will likely involve mixtures of neon and deuterium. Future simulations should investigate these mixtures to determine the composition that best balances the suppression of RE production with the mitigation of other critical disruption loads, such as thermal fluxes and halo currents. Parallel to this, work is currently underway to simulate a realistic TQ triggered by shattered pellet injection using 3D modeling, which will allow us to bypass the ATQ setup by providing more accurate initial conditions for the start of the CQ. We will also extend our analysis to 3D simulations of macroscopic RE beam termination events. This transition will simultaneously reveal how realistic MHD perturbations affect net avalanche growth compared to simplified 2D predictions and enable precise assessments of localized PFC damage, directly informing mitigation strategies like the placement of sacrificial limiters. These contributions will lay the foundation for safe DTT operations, safeguarding its strategic function within the European fusion program.

\section*{Data availability}
The data that supports the findings of this study
are available from the corresponding author upon
reasonable request.

\begin{acknowledgments}
This work has been carried out within the framework of the EUROfusion Consortium, funded by the European Union via the Euratom Research and Training Programme (Grant Agreement No 101052200 — EUROfusion). Views and opinions expressed are however those of the author(s) only and do not necessarily reflect those of the European Union or the European Commission. Neither the European Union nor the European Commission can be held responsible for them.
The simulations were performed on the Marconi-Fusion and Pitagora HPCs hosted at CINECA.
\end{acknowledgments}

\bibliography{References_Emanuelli}

\end{document}